\documentstyle[12pt]{article}
\def\mb#1{\mbox{\boldmath $#1$}}

\begin{document}

\begin{center}
{\large\bf
SPHERICAL FUNCTIONS FOR
\\[0.5em]
THE QUANTUM GROUP su\mb{_q(2)}
\footnote{Invited talk presented by P.
Winternitz at International  Workshop
``Symmetry Methods in Physics'' in memory
of Ya.~A. Smorodinsky (Dubna, Russia, July 1993).}}
\\[13pt]
{\Large P.~Winternitz}
\\[10pt]
{\sl Centre de recherches math\'ematiques, Universit\'e de
Montr\'eal}\\
{\sl C. P. 6128, Succ.~A, Montr\'eal (Qu\'ebec) Canada H3C 3J7}
\\[6pt]
{and}
\\[6pt]
{\Large G.~Rideau}
\\[10pt]
{\sl Laboratoire de Physique Th\'eorique et Math\'ematique,
Universit\'e Paris VII}\\
{\sl Tour Centrale, 2 place Jussieu, 75251 Paris Cedex 05,
France}
\end{center}
\vspace{10pt}

\begin{abstract}
The representation theory of the quantum group su$_q(2)$ is used to
introduce $q$-analogues of the Wigner rotation matrices, spherical functions,
and Legendre polynomials. The method amounts to an extension of variable
separation from Laplace equations to certain differential-dilation equations.
\end{abstract}
\vspace{1.3em}
{\bf 1. Introduction}
\medskip

One of the present authors (P.~W.) had the privilege of working for
his Ph.~D. degree under the guidance of Professor Yakov Abramovich
Smorodinsky in Dubna, starting 30 years ago. The first article he was given
to read was one by N.~Ya.\ Vilenkin and Ya.~A. Smorodinsky on invariant
expansions of relativistic scattering am\-pli\-tudes [1]. This paper
introduced a variety of different types of ``spherical functions'' as basis
functions for irreducible representations of the Lorentz group, realized on
$O(3,1)$ hyperboloids. This turned out to be a pathbreaking article that lead
to many interesting developments in the representation theory of noncompact
groups and their physical applications.

In addition to the original idea of two-variable expansion of scattering
am\-pli\-tudes [1-5], this includes a group theoretical
approach to the separation of variables in partial differential
equations [2,6-9]. Other scientific programs
influenced by Ref.~1 are those of systematically classifying subgroups of Lie
groups [2,10,11], or of generating completely integrable Hamiltonian systems
in various spaces [2,12-14].

A much more recent development is the current interest in quantum
groups [15-19] With it came the realization that the so
called $q$-hypergeometric series and other $q$-special
functions have a similar relation to quantum
groups [20-24] as the classical special functions have to
Lie groups [1,25,26].

The purpose of this presentation and an accompanying article [27] is to
use the theory of irreducible representations of the quantum group
su$_q(2)$ to construct $q$-spherical functions on an ordinary (commutative)
sphere $S^2$, i.e.\ the homogeneous space $SU(2)/O(2)$.

The motivation for this study is two-fold. First of all, if quantum groups are
to play a role in physics, then the corresponding $q$-special functions
should occur in physics as wave functions, or in some similar guise.
Secondly, a systematic use of quantum group representation theory should
provide methods for introducing new special functions and obtaining new
properties of known functions

We hope that this article demonstrates how strong the influence of one
aspect of Ya.~A. Smorodinsky's work is on research being conducted now,
almost 30 years later.
\\[1.3em]
{\bf 2. The Quantum Algebra su$_q(2)$ and its Representations}
\medskip \\
{\it 2.1 Realization of the Quantum Algebra by
Differential-Dilation Operators}
\medskip

The algebra su$_q(2)$ is a deformation of the Lie algebra su(2) and is
characterized by the commutation relations
$$
[H_3, H_+]=H_+,\quad [H_3,H_-]=-H_-,\quad
[H_+,H_-]={q^{2H_3}-q^{-2H_3}\over q-q^{-1}},\eqno(1)
$$
where $q$ is some real number. For $q\to 1$ Eq.~(1) reduce to the usual
su(2) commutation relations
$$
[H_3, H_+]=H_+,\quad [H_3,H_-]=-H_-,\quad [H_+,H_-]= 2H_3.\eqno(2)
$$
Finite-dimensional irreducible  representations of su$_q(2)$ are
characterized by an integer or half-integer number $J$. Basis functions for
these representations can be denoted $\left|JMq\right>$ and satisfy
$$
\begin{array}{l}
H_3\left|JMq\right> = M\left|JMq\right> \\ \\
H_+\left|JMq\right> = \alpha^J_{M+1,q}\left|JM+1q\right> \\ \\
H_-\left|JMq\right> = \alpha^J_{Mq}\left|JM-1q\right>
\end{array}
\eqno(3)
$$
with
$$
\alpha^J_{Mq}=\Biggl[{q^{J-M+1}-q^{-J+M-1}\over q-q^{-1}}\,
{q^{J+M}-q^{-J-M}\over q-q^{-1}}\Biggr]^{1/2}.
\eqno(4)
$$
For $q=1$ Eqs.~(3) and (4) reduce to standard $su(2)$ formulas with
$$
\alpha^J_{M1}=\alpha^J_M=[(J+M)(J-M+1)]^{1/2}.\eqno(5)
$$
Thus, in Eq.~(3)
 we have chosen a basis of eigenfunctions of the operator $H_3$,
corresponding to a nondeformed $U(1)$ subalgebra of $su_q(2)$. We have
$\alpha^J_{J+1,q}=\alpha^J_{-J,q}=0$, hence there is a highest and lowest
weight $N=\pm J$ and the representations are finite-dimensional.

We shall now pursue an analogy with the construction of spherical functions
$Y_{JM}(\theta,\phi)$ for $su(2)$. These functions can be viewed as being
defined on a sphere $S_2$ defined by the relations:
$$
x_0={1\over 2}\sin\theta\cos\phi,\quad y_0={1\over
2}\sin\theta\sin\phi,\quad z_0={1\over 2}\cos\theta.\eqno(6)
$$
Using a stereographic projection $S^2\to R^2$
$$
x={x_0\over 1/2 -z_0},\quad y={y_0\over 1/2 -z_0},\eqno(7)
$$
we can see the spherical functions as being defined on the real plane $(x,y)$
with
$$
\begin{array}{lll}
x=\rho\cos\phi,\quad y=\rho\sin\phi,\quad
\rho=\cot{\displaystyle \theta\over 2}\\ \\
0\le\theta\le \pi,\quad 0\le\theta<2\pi,\quad 0\le \rho<\infty.
\end{array}
\eqno(8)
$$
We shall also use the complex variable
$$z=x+iy=\rho e^{i\phi}.\eqno(9)$$

We now need to construct operators $H_3$, $H_+$, and $H_-$, acting on
functions $f(\theta,\phi)$ on $S_2$, or equivalently, on functions $f(z,\bar
z)$ on the plane $(z,\bar z)$. The operators should satisfy the relations~(2).

An ``inspired guess'' yields the following relations
$$
\begin{array}{l}
\displaystyle
H_3 = -z\partial_z+\bar z\partial_{\bar z}-N\\ \\
\displaystyle
H_+ = -{1\over z}{q^{z\partial_z}-q^{-z\partial_z}\over q-q^{-1}} q^{\bar
z\partial_{\bar z}-N/2} - q^{z\partial_z+N/2} \bar z {q^{\bar z
\partial_{\bar z}-N} - q^{-\bar z\partial_{\bar z}+N}\over q-q^{-1}} \\ \\
\displaystyle
H_- = z{q^{z\partial_z+N}-q^{-z\partial_z-N}\over q-q^{-1}} q^{\bar
z\partial_{\bar z}-N/2} + q^{z\partial_z+N/2} {1\over \bar z}{q^{\bar
z\partial_{\bar z}}-q^{-\bar z\partial_{\bar z}}\over q-q^{-1}},
\end{array}
\eqno(10)
$$
where $N$ is an integer or half integer parameter. The operators
$q^{z\partial_z}$ and $q^{\bar z\partial_{\bar z}}$ act like dilations:
$$
q^{z\partial_z} f(z,\bar z) = f(qz,\bar z)
$$
$$
q^{\bar z\partial_{\bar z}} f(z,\bar z) = f(z,q\bar z)
\eqno(11)
$$
and it  is easy to verify that $H_\pm$ and $H_3$ satisfy the commutation
relations~(1). For $q=1$ relations~(10) reduce to $su(2)$ expressions
$$
H_3=-z\partial_z+\bar z\partial_{\bar z}-N,\quad H_+=- \partial_z-\bar
z^2\partial_{\bar z}+N\bar z,\quad H_- =z^2\partial_z+ \partial_{\bar z}+N z.
\eqno(12)
$$

For $q=1$ we see that $H_\pm$, as well as $H_3$ are first order differential
operators. For $q\ne 1$ the operators become nonlocal: in addition to
derivatives, they involve dilations of the independent variables.
\\[1.3em]
{\it 2.2 Basis Functions for Irreducible
Representations}
\medskip

We will now look for a realization of the basis functions
$$
\left|JMq\right>=\Psi_{MNq}^J(z,\bar z)\eqno(13)
$$
satisfying Eq.~(3) with $H_\mu$ as in Eq.~(10). The Casimir operator of
$su_q(2)$, commuting with $H_\mu$, $\mu=3$, $\pm$, is
$$
C_q=H_+H_-+\Biggl({q^{H_3-1/2}-q^{-H_3+1/2}\over q-q^{-1}}\Biggr)^2-{1\over
4}.\eqno(14)
$$
 From Eq.~(3) we deduce that the basis functions of an irreducible
representation satisfy
$$
C_q\Psi_{MNq}^J=\Biggl[\Biggl({q^{J+1/2}-q^{-J-1/2}\over
q-q^{-1}}\Biggr)^2-{1\over 4}\Biggr]\Psi_{MNq}^J.\eqno(15)
$$
For $q\to 1$ Eq.~(15) reduces to the standard angular momentum relation
$$
C\Psi_{MN}^J=J(J+1)\Psi_{MN}^J.\eqno(16)
$$
We are after explicit expressions for the basis functions~(13) that for $q=1$,
$N=0$ reduce to $su(2)$ spherical functions (and for $N\ne 0$ to Jacobi
polynomials). To do this we put:
$$
\Psi_{MNq}^J(z,\bar z) =N_{MNq}^Jq^{-NM/2}Q_{Jq}(\eta)
R_{NMq}^J(\eta)\bar z^{M+N},
\eqno(17)
$$
$$
\eta =z\bar z=\cot^2\Biggl({\theta\over 2}\Biggr).
$$
To get a finite-dimensional representation (of dimension $2J+1$) we request
the existence of a highest and lowest weight, i.e.
$$
H_+\Psi_{JNq}^J=0,\quad H_-\Psi_{-JNq}^J=0\eqno(18)
$$
and by analogy with the $q=1$ case, we put
$$
R_{JNq}^J=\hbox{const}.\eqno(19)
$$
The first of relations~(18) implies a functional relation for $Q_{Jq}(\eta)$,
namely
$$
Q_{Jq}(q^2\eta)(1+\eta)=Q_{Jq}(\eta)(1+q^{-2J}\eta).\eqno(20)
$$
Its solution in terms of Exton's $q$-binomial function$^{28}$ is
$$
Q_{Jq}(\eta)={}_1\phi_0(J;-;q^2;-\eta q^{-2J})\eqno(21)
$$
which for $q\to 1$ reduces to $Q_{J1}(\eta)=(1+\eta)^{-J}$.

For $q=1$ the expression $R_{MN1}^J(\eta)$ are polynomials related to the
Jacobi polynomials. For general $q$ they are also polynomials, satisfying
certain relations, following from Eq.~(3). They are
$$
(\alpha_{M+1,q}^J)^2\eta R_{M+1,N,q}^J(\eta)
={q^{M+N}\over q-q^{-1}}\{-(1+q^{-2J}\eta) R_{MNq}^J(q^2\eta)
+(1+q^{-2M}\eta) R_{MNq}^J(\eta)\}
\eqno(22)
$$
$$
R_{M-1,N,q}^J(\eta) = {1\over q-q^{-1}}\{q^{M+N}(1+q^{-2J}\eta)
R_{MNq}^J(q^2\eta)
-q^{-M-N}(1+q^{2M}\eta) R_{MNq}^J(\eta)\}
\eqno(23)
$$
The normalization constant $N_{MNq}^J$ in Eq.~(17) was chosen to be
$$
N_{MNq}^J=C_{JNq}\Biggl({[J+M,q]!
\over [J-M,q]![2J,q]!}\Biggr)^{1/2},\eqno(24)
$$
where $[a,q]!$ denotes a $q$-factorial and $[a,q]$ a $q$-number$^{28,29}$
$$
[a,q]={q^a-q^{-a}
\over q-q^{-1}},\quad [k,q]!=\prod_{p=1}^k[p,q],
$$
$$
[0,q]!=1,\quad {1\over [k,q]!}=0,\quad k\in Z^{<0}.
\eqno(25)
$$

Relations~(22) and (23) are nonlocal, in that the functions $R_{MNq}^J$ are
evaluated at $\eta$ and $q^2\eta$, We can eliminate $R_{MNq}^J(q^2\eta)$ from
these two ``difference-dilation'' equations and obtain a recursion relation,
namely
$$
(\alpha_{M+1,q}^J)^2\eta R_{M+1,N,q}^J(\eta)+ R_{M-1,N,q}^J(\eta)=\{[M+N,
q]- [M-N,q]\eta\}R_{MNq}^J(\eta).\eqno(26)
$$
\vspace{1.3em} \\
{\bf 3. The \mb{q}-Vilenkin-Wigner Functions and $q$-Spherical
Functions}
\medskip

The recursion relation~(26), as well as the relations~(22) and (23) are solved
by the following polynomial in $\eta$:
$$
\begin{array}{l}
R_{MNq}^J(\eta)=[J-N,q]![J-M,q]!\times \\
\qquad\qquad\quad \displaystyle
\times \sum_k {(-1)^k\eta^k \over
[k,q]![J-M-k,q]! [J-N-k,q]![M+N+k,q]!}
\end{array}
\eqno(27)
$$
and thus the basis functions~(17) are completely determined. With an
appropriate choice of the normalization constant $C_{JNq}$ in Eq.~(24) we can
rewrite them as
$$
\Psi_{MNq}^J(\theta,\phi)={1\over \sqrt{2\pi}}([2J+1,q])^{1/2}i^{-2J+M+N}
P_{MNq}^J(\cos\theta)e^{-i(M+N)\phi}.\eqno(28)
$$
Here we have introduced the ``$q$-Vilenkin-Wigner functions''
$$
P_{MNq}^J(\xi)=i^{2J-M-N} \Biggl({[J+M,q]![J+N,q]!\over
[J-M,q]![J-N,q]!}\Biggr)^{1/2}\eta^{(M+N)/2}Q_{Jq}(\eta)R_{MNq}^J(\eta)
$$
$$
\eta=z\bar z={1+\xi\over 1-\xi},\quad \xi=\cos\theta.
\eqno(29)
$$
For $q=1$ these functions reduce to the functions $P_{MN}^J(\cos\theta)$
extensively studied by Vilenkin$^{25}$ and directly related to the Wigner
rotation matrices$^{30}$ $d_{MN}^J(\theta)$. They are also related to
Jacobi polynomials
$$
\begin{array}{l}
\displaystyle
P_{MN}^J(\xi)=2^{-M}(i)^{-N+M}\Biggl[{(J-M)!(J+M)!\over
(J-N)!(J+N)!}\Biggr]^{1/2} \times \\
\qquad\qquad\  \times (1-\xi)^{(-N+M)/2}(1+\xi)^{(N+M)/2}
P_k^{(p,q)}(\xi) \\ \\
\qquad k=J-M,\quad p=M-N,\quad q=M+N.
\end{array}
\eqno(30)
$$
The functions $P_{MN}^J(\xi)$ are usually introduced as rotation matrices,
but they can just as well arise as basis functions for representations of
$su(2)$. It is this second role, that of basis functions for irreducible
representations, that has been generalized to the quantum group $su_q(2)$.

Most properties of the ordinary ($q=1$) Vilenkin-Wigner functions have their
$q$-analogues. We shall just list some of them and refer to a related
article$^{27}$ for proofs

\medskip
\noindent{\it Recursion formula}
$$
\begin{array}{l}
i([M+N,q]\eta^{-1/2}-[M-N,q]\eta^{1/2}) P_{MNQ}^J = \\ \\
\qquad\qquad\quad = -([J-M,q][J+M+1,q])^{1/2} P_{M+1,N,q}^J + \\  \\
\qquad\qquad\qquad\quad + ([J-M+1,q][J+M,q])^{1/2} P_{M-1,N,q}^J.
\end{array}
\eqno(31)
$$

\noindent{\it Generating function}, defined as
$$
F_{N,q}^J(\omega)=\sum_{M=-J}^J{P_{MNq}^J(\eta)\over
([J-M,q]![J+M,q]!)^{1/2}}\omega^{J-M},
\eqno(32)
$$
has the form
$$
F_{N,q}^J(\omega)= i^{J+N}{1\over ([J+N,q]![J-N,q]!)^{1/2}}
\eta^{(J-N)/2}Q_{Jq}(\eta)
\qquad\qquad
$$
$$
\qquad\qquad\qquad \prod_{p=0}^{J-N-1}(\omega q^{-J+N+1+2p}+i\eta^{-1/2})
\prod_{p=0}^{J+N-1}(\omega q^{-J-N+1+2p}-i\eta^{1/2}).
\eqno(33)
$$
For $q=1$ this simplifies to the well-known generating function
$$
F_{N1}^J(\omega)\equiv F(\omega)={1\over
[(J-N)!(J+N)!]^{1/2}}\biggl(\cos{\theta
\over 2}+i\omega\sin{\theta\over 2}\biggr)^{J+N} \times
$$
$$
\times\biggl(\omega\cos{\theta
\over 2}+i\sin{\theta\over 2}\biggr)^{J-N}.
\eqno(34)
$$

{\it Symmetry relations}
$$
P_{MNq}^J(\xi) = P_{NMq}^J(\xi) \eqno(35)
$$
$$
P_{MNq}^J(\xi) = P_{-M,-N,q}(\xi) \eqno(36)
$$
$$
P_{MNq}^J(-\xi) = i^{2J-2M-2N}\sigma_{J,q}{Q_{Jq}(\eta q^{2J+2})\over
Q_{Jq}(\eta)} P_{M,-N,q}(\xi) \eqno(37)
$$
where
$$
\begin{array}{ll}
\sigma_{J,q}=
q^{J(J+1)}\phantom{\sigma_{1/2,q} \hbox{ffor }}
& \hbox{for $J$ integer}\cr \\
\sigma_{L+1/2,q} =
q^{(L+1)^2}\sigma_{1/2,q} & \hbox{for $J$ half-odd integer}\cr \\
\displaystyle
\sigma_{1/2,q}={1+q\over 1+q^{-1}}{1\over\sqrt{q}}{\theta_2(0)\over
\theta_3(0)},
\end{array}
$$
($\theta_2(u)$ and $\theta_3(u)$ are ordinary theta functions).

It is now quite natural to introduce $q$-spherical harmonics in the same
manner as ordinary harmonics, namely
$$
Y_{JMq}(\theta,\phi)={1\over
[J-M,q]!}P_{M0q}^J(\cos\theta)e^{-iM\phi}.\eqno(38)$$
Similarly, the $q$-analogue of Legendre polynomials is
$$P_{Jq}(\cos\theta)= P_{00q}^J(\cos\theta)= i^{2J}Q_{Jq}(\eta)
R_{00q}^J(\eta).\eqno(39)
$$
Notice that $P_{Jq}(\cos\theta)$ for $q\ne 1$ is not a polynomial in
$\cos\theta$ in view of the properties of $Q_{Jq}(\eta)$.

Finally we mention the relations between the $q$-Vilenkin-Wigner functions
and other $q$-functions in the literature. These relations are best written
in terms of the polynomials $R_{MNq}^j(\eta)$ of Eq.~(27). For instance, in
terms of the basic hypergeometric series$^{28,29}$
$$
{}_2F_1(a,b;c;q;z)=\sum_{k=0}^\infty {[a,k;q][b,k;q]\over
[c,k;q][k,q]!}z^k
\eqno(40)
$$
$$
[a,k;q]=[a,q][a+1,q]\cdots[a+k-1,q]
$$
we have, for $M+N\ge 0$
$$
R_{MNq}^J(\eta)=
{1\over[M+N,q]!}\,{}_2F_1(M-J,N-J;M+N+1;q;-\eta).\eqno(41)
$$

Using the ``little $q$-Jacobi functions'' $p_n(x;a,b;\eta)$ given e.g.\ by
Koornwinder$^{31}$, we have
$$
R_{MNq}^J={1\over[M+N,q]!}P_{J-M}(-q^{2j-1}\eta;q^{2(M+N)},q^{-2(2J+1)};
q^2)\eqno(42)
$$
(for $J-M,J-N$).
\\[1.3em]
{\bf 4. Conclusions}
\medskip

The full power of Lie group theory in its application to special functions
only becomes apparent, when applied to partial differential equations and
combined with the separation of variables. One way of viewing the results
presented above is that we have extended the Lie algebraic treatment of
variable separation to $q$-special functions and to quantum groups. The
separation occurs in differential-difference equations of type~(14)
and~(15), rather than in Laplace-Beltrami equations.

In the future we plan to apply similar techniques to other quantum groups and
hence to other types of $q$-special functions.
\\[1.3em]
{\bf Acknowledgements}
\medskip

The research of P. W. was partially supported by grants from NSERC of
Canada and FCAR du Qu\'ebec.


\vspace{1cm}
\begin{thebibliography}{99}
\bibitem{1}
N.~Ya.\ Vilenkin and Y.~A. Smorodinsky, {\it Zh.\ Eksp.\
Teor.\ Fiz}.~{\bf 46} (1964) 1793.
\bibitem{2}
P.~Winternitz and I.~Fri\v s, {\it Yad.\ Fiz}.~{\bf 1}
(1965) 889.
\bibitem{3}
P. Winternitz, Y.~A. Smorodinsky, and M.~B. Sheftel, {\it
Yad.\ Fiz}.~{\bf 7} (1968) 1325.
\bibitem{4}
M.~Daumens and P.~Winternitz, {\it Phys.\ Rev}.~{\bf
D21} (1980) 1919.
\bibitem{5}
J.~Bystricky, P.~LaFrance, F.~Lehar, F.~Perrot, and P.~Winternitz,
{\it Phys.\ Rev}.~{\bf D32} (1985) 575.
\bibitem{6}
E.~G. Kalnins, W.~Miller Jr., and P.~Winternitz,
{\it SIAM J. Appl}.~{\bf 30} (1976) 630.
\bibitem{7}
W.~Miller Jr., J.~Patera, and P.~Winternitz, {\it
J. Math.\ Phys}.~{\bf 22} (1981) 251.
\bibitem{8}
W.~Miller Jr., {\it Symmetry and Separation of Variables}
(Addison Wesley, New York, 1977).
\bibitem{9}
E.~G. Kalnins, {\it Separation of Variables for Riemannian
Symmetric Spaces of Constant Curvature} (Longmans, Essex, England, 1986).
\bibitem{10}
J.~Patera, P.~Winternitz, and H.~Zassenhaus, {\it J. Math.\
Phys}.~{\bf 15} (1974) 1378 and 1932; {\bf 16} (1975) 1597 and
1615; {\bf 17} (1976) 717.
\bibitem{11}
J.~Patera, R.~T. Sharp, P.~Winternitz, and H.~Zassenhaus,
{\it J. Math.\ Phys}.~{\bf 17} (1976) 977 and 986; {\bf 18}
(1977) 2259.
\bibitem{12}
I.~Fri\v s, V.~Mandrosov, J.~Smorodinsky, M.~Uhli\v r, and
P.~Winternitz, {\it Phys.\ Lett}.~{\bf 16} (1965) 354.
\bibitem{13}
A.~Makarov, J.~Smorodinsky, Kh.~Valiev, and P.~Winternitz,
{\it Nuovo Cim}.~{\bf A52} (1967) 1061
\bibitem{14}
M.~A. del~Olmo, M.~A. Rodriguez, and P.~Winternitz,
{\it J. Math.\ Phys}.~{\bf 34} (1993) 5118.
\bibitem{15}
V.~G. Drinfeld, in {\it Proc.\ Int.\ Congress Math}.~Vol 1
(Amer.\ Math.\ Soc., Providence, RI, 1986).
\bibitem{16}
M.~Jimbo, {\it Lett.\ Math.\ Phys}.~{\bf 10} (1985)
63; {\bf 11} (1986) 247.
\bibitem{17}
S.~L. Woronwicz,
{\it Commun.\ Math.\ Phys}.~{\bf 111}
(1987) 613.
\bibitem{18}
Yu.~I. Manin, {\it Quantum Groups and Noncommutative
Geometry} (Centre de recherches math\'ematiques, Montr\'eal, 1988).
\bibitem{19}
L.~D. Faddeev, N.~Y. Reshetikhin, and L.~A. Taktajan,
{\it Leningrad.\ Math.\ J}.~{\bf 1} (1990) 193.
\bibitem{20}
R.~Floreanini and L.~Vinet, {\it Lett.\ Math.\
Phys}.~{\bf 22} (1991) 45; {\it J. Phys. A Math.\
Gen}.~{\bf 23} (1990) L1019; {\it J. Math.\ Phys}.~{\bf
33} (1992) 1358.
\bibitem{21}
E.~G. Kalnins, H.~L. Manocha, and W. Miller Jr.,
{\it J. Math.\ Phys}.~{\bf 33}, (1992) 2365.
\bibitem{22}
L.~L. Vaksman and Ya.~S. Soibelman, {\it Funct.\ Anal.\
Pril}.~{\bf 22} (1988) 1.
\bibitem{23}
N. M. Atakishiyev and S.~K. Suslov, {\it Teor.\
Mat.\ Fiz}.~{\bf 85} (1990) 64.
\bibitem{24}
S.~K. Suslov, {\it Russian Math.\ Surveys}~{\bf 44}
(1989) 227.
\bibitem{25}
N.~Ya. Vilenkin, {\it Special Functions and the Theory of
Group Representations} (Amer.\ Math.\ Soc., Providence, RI, 1968).
\bibitem{26}
W.~Miller Jr., {\it Lie Theory and Special Functions}
(Academic Press, New York, 1986).
\bibitem{27}
G.~Rideau and P.Winternitz, {\it J. Math.\ Phys}.~{\bf
34} (1993) 6030.
\bibitem{28}
H.~Exton, {\it $q$-Hypergeometric Functions and
Applications} (Ellis Harwood, Chichester, 1983).
\bibitem{29}
G.~Gasper and M~Rahman, {\it Basic Hypergeometric Series}
(Cambridge Univ.\ Press, Cambridge, 1990).
\bibitem{30}
E.~P. Wigner, {\it Group Theory and its Applications to the
Quantum Mechanics of Atomic Spectra} (Academic Press, New York, 1959).
\bibitem{31}
T.~H. Koornwinder,
{\it Proc.\ Nederl.\ Akad.\ Wetensch}.
{}~{\bf A92} (1989) 97;
{\it SIAM J.\ Math.\ Anal}.
{}~{\bf 22} (1991) 295.
\end{thebibliography}
\end{document}